\documentstyle[12pt,epsf]{article}

\input psfig
\newcount\notenumber
\def\clearntnumber{\notenumber=0}
\def\nt{\advance\notenumber by1 \footnote{$^{\the\notenumber}$}}
\clearntnumber

\def\simlt{\hbox{ \rlap{\raise 0.425ex\hbox{$<$}}\lower 0.65ex
  \hbox{$\sim$} }}
\def\simgt{\hbox{ \rlap{\raise 0.425ex\hbox{$>$}}\lower 0.65ex
  \hbox{$\sim$} }}

\font\lb=cmbx10 scaled \magstep1

\font\lr=cmr10 scaled \magstep1

\def\part#1{\vfill\eject\hbox{}\bigskip\bigskip
  {\parindent=0pt\lb #1}\bigskip}

\def\chapter#1{\hbox{}\bigskip\goodbreak
  {\parindent=0pt\bf #1}}
\def\section#1{\hbox{}\goodbreak
  {\parindent=0pt\lr #1}}

\def\ref{{\parindent=0pt\medskip\hangindent=3pc\hangafter=1}}
\title{Astrophysics on the GRAPE Family of     Special Purpose Computers}

\author{Piet Hut\\
Institute for Advanced Study\\
Princeton, NJ 08540, U.S.A.\\
\\
Jun Makino\\
Department of General Systems Study\\
College of Arts and Sciences, University of Tokyo\\
3-8-1 Komaba, Meguro-ku, Tokyo 153-8902, Japan
}
\begin{document}

\maketitle

\begin{abstract}
The GRAPE-4, the world's fastest computer in 1995-1997, has produced
some major scientific results, through a wide diversity of large-scale
simulations in astrophysics.  Applications have ranged from planetary
formation, through the evolution of star clusters and galactic nuclei,
to the formation of galaxies and clusters of galaxies.
\end{abstract}

\newpage

\vfill\eject
     

Computational physics has emerged as a third branch of physics,
grafted onto the traditional pair of theoretical and experimental
physics.  At first, computer use seemed to be a straightforward
off-shoot of theoretical physics, providing solutions to sets of
differential equations too complicated to solve by hand.  But soon the
quantitative improvement in speed yielded a qualitative shift
in the nature of these computations.  Rather than asking particular
questions about a model system, we now use computers more often to
model the whole system directly.  Answers to relevant questions are
then extracted only after a full simulation has been completed.  The
data analysis following such a virtual lab experiment is carried out
by the computational physicist in much the same way as it would be
done by an experimenter or observer analyzing data from a real
experiment or observation.

Recent increase in computer speed is already significantly more modest
than what could be expected purely from the ongoing miniaturization of
computer chips.  Since the number of transistors on a single chips
doubles every $1.5$ years, a chip now contains a hundred times more
transistors than it did ten years ago.  With a clock speed increase of
more than a factor ten, one might have expected a speed increase of
more than a factor thousand, over the last decade.  However, the
actual speed increase of a typical computer chip has been at most a
factor hundred, lagging far behind theoretical expectations.  The
reason for this relatively poor performance lies in the significant
overhead caused by the growing complexity of a general-purpose chip.
Hence, designing a chip for only one specific purpose yields a
rapidly growing pay-off.  Therefore, the time seems ripe to explore
which types of calculations can be realized directly in hardware, in
the form of special-purpose computers, rather than run in software on
general-purpose computers.

One of these projects has resulted in the GRAPE (short for GRAvity
PipE) family of special-purpose hardware, designed and built by a
small group of astrophysicists at the University of Tokyo
\cite{MakinoTaiji1998}.  Like a graphics accelerator speeding up
graphics calculations on a workstation, without changing the software
running on that workstation, the GRAPE acts as a Newtonian force
accelerator, in the form of an attached piece of hardware.  In a
large-scale gravitational $N$-body calculation, where $N$ is the
number of particles, almost all instructions of the corresponding
computer program are thus performed on a standard workstation, while
only the gravitational force calculations, in innermost loop, are
replaced by a function call to the special-purpose hardware.

Specifically, the force integration and particle pushing are all done
on the host computer, and only the inter-particle force calculations
are done on the GRAPE (fig. 1).  This may seem problematic, given the
fact that the intrinsic speed of the GRAPE is a factor of 10,000 times
larger than that of the host computer, an ordinary workstation.
However, the inter-particle calculations require a computer processing
power that scales with $N^2$, while all other actions on the host
scale only in proportion to $N$.  Therefore, each doubling of the
number of particles doubles the work load on the GRAPE, relative to
that of the workstation.  In this way, no matter how slow the
workstation is, it will be able to keep up with the GRAPE for large
enough values of $N$.

For some applications, more efficient algorithms have been deviced,
that require the computation of a number of inter-particle force
calculations that scales with $N\log N$, rather than $N^2$.  It turns
out that even these methods can still be efficiently run on the GRAPE
\cite{treecodes}; although the asymptotic scaling advantage is not
very large in that case, the overall coefficient in the scaling
relation turns out to favor the use of the GRAPE.  Some versions of
the GRAPE (Table 1) allow arbitrary force implementations, for
applications such as molecular dynamics.  For example, the MDGRAPE has
been used to study the structure of protein molecules \cite{proteins}.
However, most GRAPEs have been used to study astrophysical problems.
Below we will review a few representative cases.

\begin{figure}
\begin{center}
\leavevmode
\psfig{figure=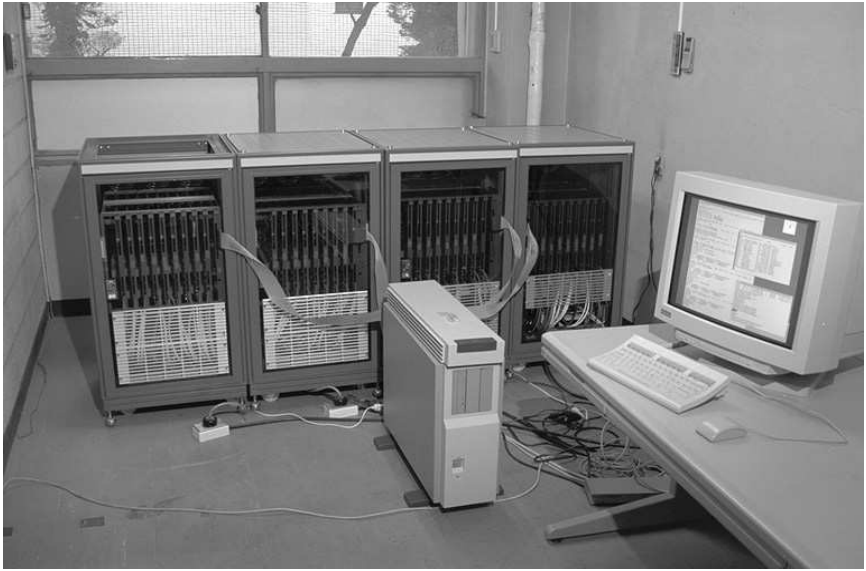,width=4.5in}
\caption{The GRAPE-4 hardware.}
\end{center}
\end{figure}

\begin{table}
\begin{center}
\caption{Summary of GRAPE Hardware (\protect\cite{MakinoTaiji1998})}
\vbox{\hbox to \hsize{\hfil\vbox{\tabskip=1 em \halign
  {#\hfil&\hfil#\hfil&\hfil#\hfil&#\hfil\cr
\noalign{\hrule}\cr
\noalign{\smallskip}
\noalign{\hrule}\cr
\noalign{\medskip}
\quad{\it Limited-Precision Data Path}\hidewidth & & & \cr
\noalign{\medskip}
Machine  & Year & Peak Speed & Notes\cr
\noalign{\smallskip}
GRAPE-1  & 1989 & 240 Mflops       & Concept system\cr
GRAPE-3  & 1991 & 15 Gflops        & 48 Custom chips, 10 MHz clock\cr
GRAPE-5  & (1999) & $\sim 1$ Tflops        & under development\cr
\noalign{\bigskip}
\noalign{\medskip}
\quad{\it Full-Precision Data Path}\hidewidth & & & \cr
\noalign{\medskip}
GRAPE-2  & 1990 & 40 Mflops        & IEEE precision, commercial chips\cr
HARP-1   & 1993 & 180 Mflops       & force and its time derivative\cr
GRAPE-4  & 1995 & 1.1 Tflops       & 1692 Custom chips, 32 MHz clock\cr
GRAPE-6  & (2000) & $\sim 200$ Tflops       & under development\cr
\noalign{\bigskip}
\noalign{\medskip}
\quad{\it Arbitrary Force Law}\hidewidth & & & \cr
\noalign{\medskip}
GRAPE-2A & 1992 & 180 Mflops       & Force look-up table\cr
MDGRAPE & 1995 & 4 Gflops         & Custom chip with force look-up table\cr
MDGRAPE-2 & (2000) & $\sim 100$ Tflops         & under development\cr
\noalign{\medskip}
\noalign{\hrule}\cr
\noalign{\medskip}}}\hfil}}
\end{center}
\end{table}

\chapter{Star Cluster Evolution} 

A globular star cluster [fig. 2; \cite{R88}] typically contains about
a million stars, packed together much closer than the stars in the
solar neighborhood.  Such a cluster describes a wide orbit around the
parent galaxy, well separated from the stars in that galaxy.  If we
would live in the core a dense globular cluster, the brightest stars
would appear as bright as the full moon, which would make them too
bright too look at directly, given their point-like nature.  Optical
astronomy of anything but the nearby stars in the same globular
cluster would be rather difficult, in such a situation.

\begin{figure}
\begin{center}
\leavevmode
\psfig{figure=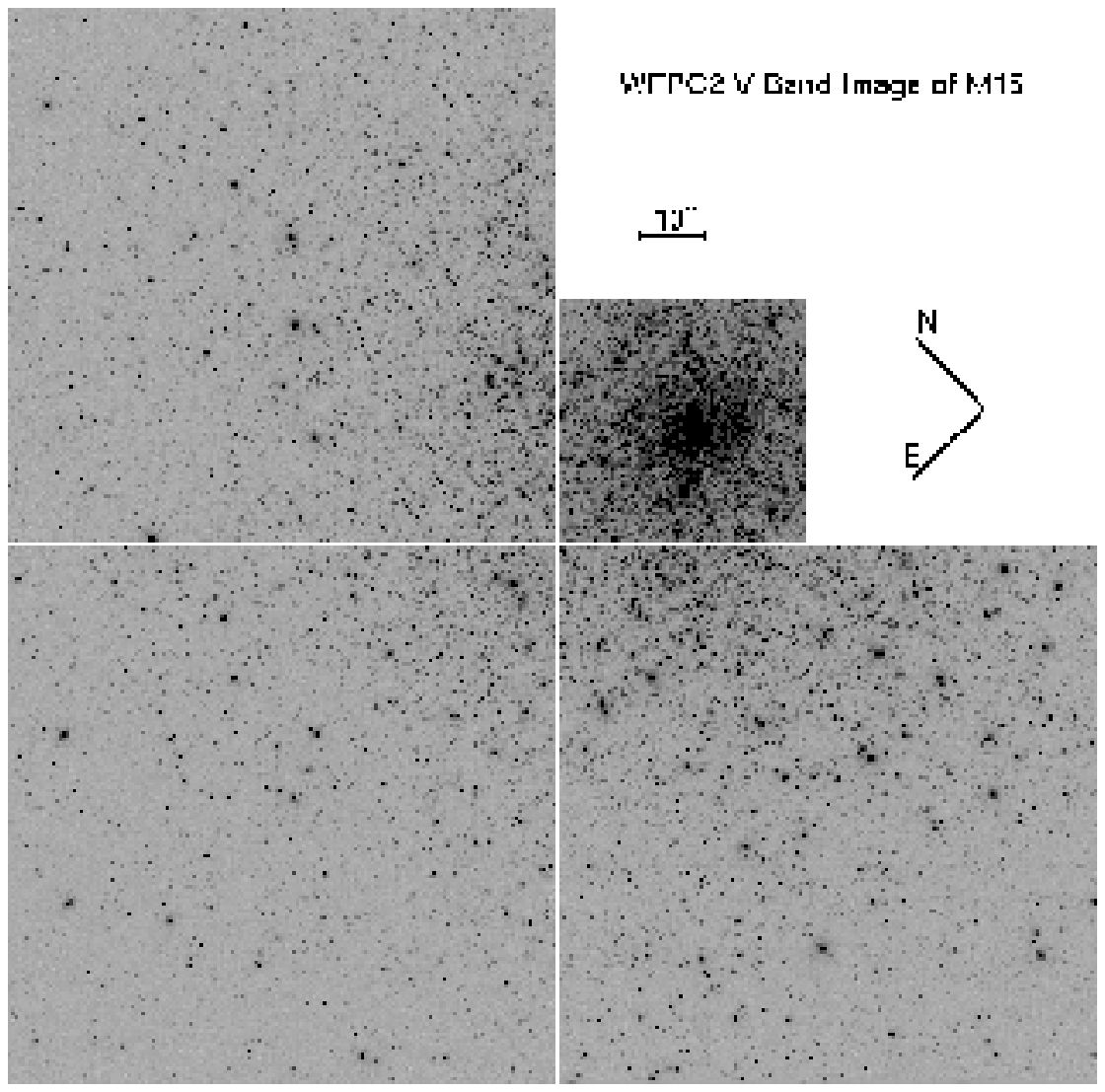,width=4in}
\end{center}
\caption{A globular cluster, M15, as seen by the Hubble Space
Telescope\protect\cite{R88}}
\end{figure}

In the simplest approximation, we can study a globular cluster as a
collection of point masses, reducing the problem to the gravitational
$N$-body problem, which was solved by Newton for $N=2$, but was only
studied in detail for $N>2$ when computers became available.  Any
localized distribution of particles will tend to become spherical, as
a result of forgetting the initial conditions, on a two-body
relaxation time scale:
$$
t_{rel} \sim 0.1 {N\over\ln N} t_{cr},
$$
where the crossing time $t_{cr}$ is a measure for the time it takes
for a typical star to move across the cluster.

Heat is transported through the cluster, as a consequence of many
two-body encounters, on the time scale $t_{rel}$.  On longer time
scales, any self-gravitating star system is unstable.  Since the
system tends to relax towards a Maxwellian velocity distribution,
there are always some stars that acquire a velocity that exceeds the
escape velocity, after which they are lost from the system.  Other
stars tend to congregate in the central regions which grow denser at
an ever-increasing rate, because higher density implies more frequent
encounters and hence a faster two-body relaxation.

This run-away redistribution of energy and mass leads to a phenomenon
called gravo\-thermal collapse, often called core collapse, which
takes place on a time scale $t_{cc} \sim 10 t_{rel}$.  Core collapse
was hinted at in numerical simulations \cite{Henon1961} in the 1960s,
and verified through direct $N$-body simulations
\cite{AarsethHenonWielen1974} and modeled by semianalytic methods
\cite{LyndenBellEggleton1980} in the 1970s.
Core collapse is a fundamental feature of long-term stellar-dynamical
evolution, showing the instability that results from the negative
specific heat inherent in self-gravitating systems.  During core
collapse, at first the system can be modeled as passing through a
series of self-gravitating equilibrium models exhibiting a maximum
entropy for a finite central concentration.  Once this maximum is
passed, subsequent evolution will increase the entropy, and the
structure of the star cluster is forced to deviate from that of an
equilibrium model.

Even in an idealized system of self-gravitating point particles, core
collapse will be halted before an infinite central density is reached.
When the central density is high enough, occasional close encounters
between three unrelated particles will form bound pairs (binary stars
in the case of star clusters), with the third particle carrying off
the excess kinetic energy required to leave the other two particles
bound.  Subsequent encounters between such pairs and other single
particles tend to increase the binding energy of these pairs, which
leads to a heating of the surrounding system of single particles.

     When enough pairs have been formed in this way, the resulting energy
production will reverse the process of core collapse.  After reaching
a minimum radius and a maximum density, the core region will expand
again.  Core collapse, when threatened to occur by the collective
effects of two-body relaxation, can thus be narrowly averted by a
handful of crucial three-body or four-body reactions in the dense
core of a nearly collapsed cluster.  What will happen next depends on
the total number $N$ of particles in the system.  If this number is
sufficiently small, $N \simlt 10,000$, the whole system will
slowly and steadily expand.  In this case a steady-state equilibrium
can be found between the steady energy production in three-body
encounters in the center, and the continuous loss of energy through
the outskirts of the system.

     If the total number of particles exceeds $10^4$, however, a
different behavior emerges.  The more particles there are in the
system, the higher the central density has to become to halt core
collapse.  As a result, the post-collapse phase features a short
relaxation time in the center of the cluster, shorter than the
relaxation time in the outer regions, where most of the particles can
be found.  From the point of view of the inner core dynamics, the bulk
of the mass further out seems almost frozen.  It is this discrepancy
in time scales that can cause the inner core to become `impatient',
and to revert to a local collapse, triggered by the slightest
fluctuation in the direction of the energy flow produced by stochastic
three- and four-body interactions.

     What happens then is that about 1\% of the inner particles will
go into a coherent collapse, locally reminiscent of the original core
collapse.  As before, bound pairs of particles spring into action,
generate energy, and manage to reverse the collapse in the nick of
time, preventing an infinite central density from building up.  This
process repeats itself, leading to irregular oscillations of the core
of the cluster.

     The existence of these oscillations was unknown until 1983, when
they were first found in approximate simulations \cite{R3}.  Dubbed
`gravothermal oscillations', they were subsequently analyzed in detail
with semi-analytic methods \cite{R4}.  Their occurrence was confirmed
in a variety of approximate numerical simulations \cite{R5}, and shown
to correspond to low-dimensional chaos for large $N$ values \cite{R6}.
Direct verification of the existence of these oscillations was
attempted, using the fastest supercomputers available, but these
attempts were unsuccessful \cite{MakinoSugimoto1987, SpurzemAarseth1996}.

The reason that they were so hard to confirm through direct $N$-body
simulations lies in the fact that it has not been possible to model
star cluster evolution with more than 10,000 particles until the
advent of the GRAPE-4.  This may seem surprising, given the fact that
cosmological simulations now routinely handle up to a billion
particles.  The main difference between the two type of calculations
lies in the higher accuracy required for star cluster simulations,
together with the much larger number of time steps required, in
comparison with cosmological simulations.

As for the first point, following the gravothermal collapse requires a
very accurate integration of the equations of motion. The required
accuracy is difficult to achieve using approximate schemes like tree
codes \cite{BarnesHut}.  Therefore, traditional direct summation schemes
have to be used.  Even on supercomputers the maximum particle number
is thus limited to about ten thousand.

The second point is related to the fact that $N$-body simulations play
a very different role in the modeling of star clusters, and of
cosmological large-scale structure formation.  In the case of star
clusters, each particle stands for an individual star, and thus has a
direct physical meaning.  In the case of a cosmological simulation,
each galaxy is represented by a relatively small number of particles,
that sample the distribution of stars in phase space.  Each particle
thus represents the average behavior of many millions of stars.  The
time steps used can therefore be much larger than would be the case if
we were to follow the close encounters of individual stars.

Finally, the existence of gravo-thermal oscillations was proven when
they were seen in a direct $N$-body simulation on the GRAPE-4 that was
able to incorporate $N$ values beyond $N=10,000$ \cite{R7}, as
illustrated in figures 3 and 4.

     After core collapse, the fluctuations in central density grow
with increasing $N$ values, as is clear from figure 3.  For the
largest $N$ values displayed, the typical behavior of core
oscillations emerges, with its deep and long-lasting throughs
punctuated with brief interludes of high core density.  For smaller
$N$ values, some oscillatory behavior seems to be present, but less
pronounced.  The results of the central density evolution, while
suggestive, do not answer the question of the existence of gravothermal
oscillations.

Figure 4, however, provides the proof of the gravothermal nature of
these oscillations \cite{R8}.  The thermodynamic cycle exhibited by
the central density and `temperature' (as measured by the velocity
dispersion), is traversed in the opposite direction from that of a
Carnot engine: the decompression stage takes place at a lower
temperature than the compression stage.  This is a reflection of the
negative heat capacity of self-gravitating systems: compression leads
to a temperature increase resulting in more heat loss and hence more
compression, with the opposite effects holding during decompression.
The period of decompression finishes when the core expands beyond the
central isothermal area.

\begin{figure}
\begin{center}
\leavevmode
\psfig{figure=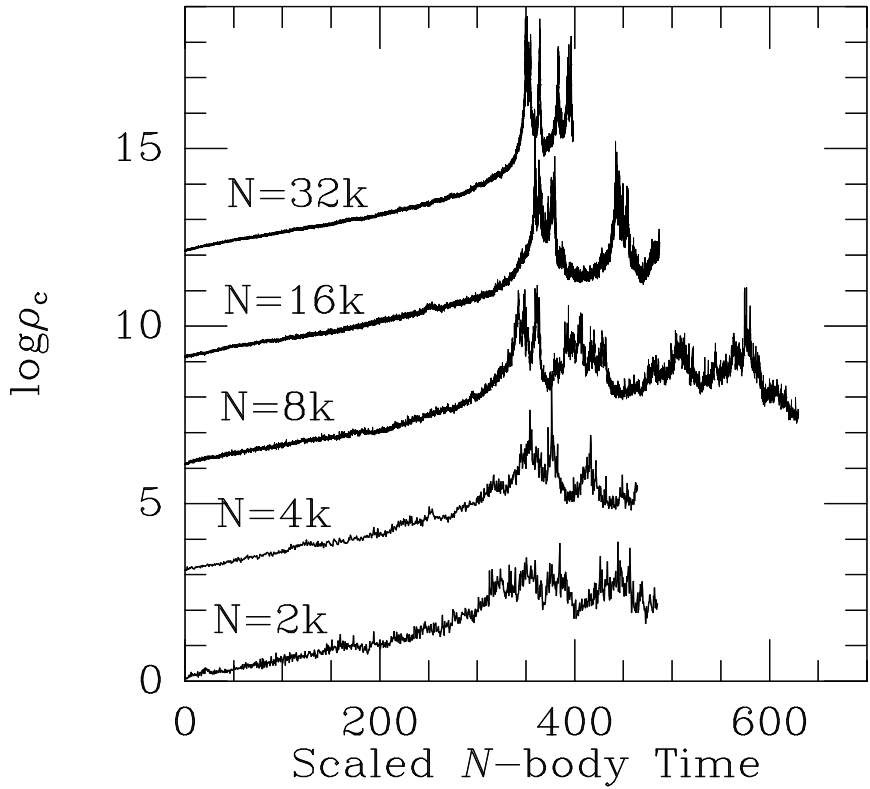,width=4.5in}
\caption{The evolution of the central density $\rho_c$.  Thirty time units
correspond roughly to one initial half-mass relaxation time.
Curves for different values of $N$ are vertically
shifted by 3 units.}
\end{center}
\end{figure}

\begin{figure}
\begin{center}
\leavevmode
\psfig{figure=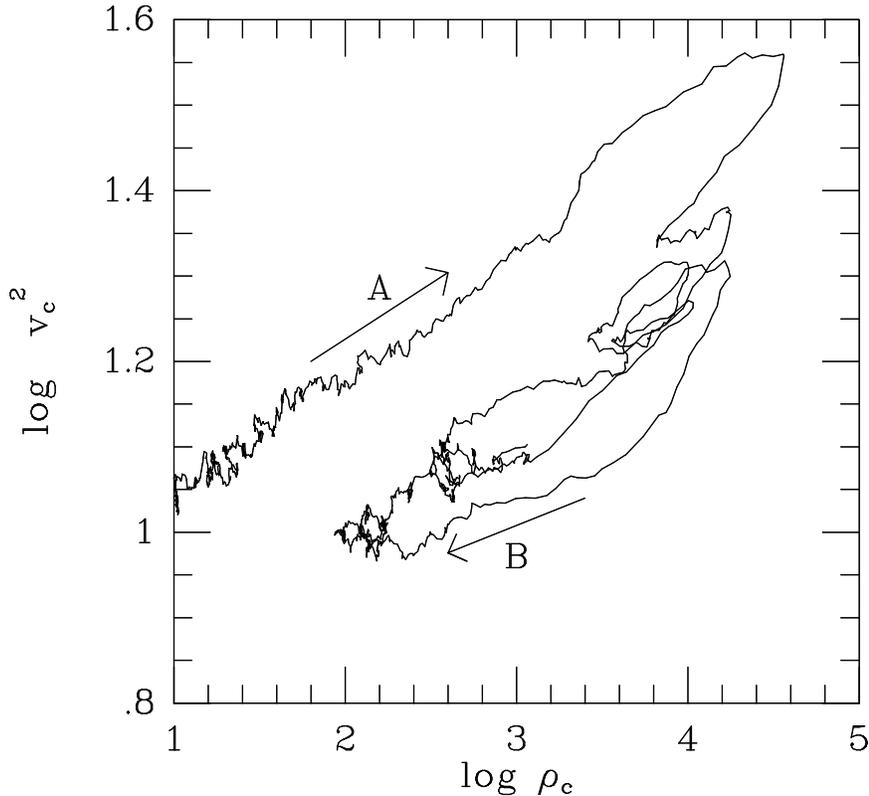,width=4.5in}

\caption{Changes in central density $\rho$ and central velocity
dispersion $v_c^2$, for a simulation with 32k particles.  Each data
point presents a time average, obtained by averaging $\rho$ and
$v_c^2$ over 80 snapshots. Arrows indicate the direction of
evolution.}
\end{center}
\end{figure}

Having clarified the fundamental behavior of self-gravitating point
mass systems, in the limit of very large numbers of particles, we are
currently working on more realistic treatments of star clusters, where
the evolution of individual stars is modeled \cite{R9}.

\chapter{Black Hole Spiral-In}

When two galaxies collide, they are likely to stick together, if their
relative speed is not too high.  Within a few crossing times, the
transient ripples and distortions will be smoothed out, and the
resulting single galaxy will settle down into a new equilibrium
configuration.  While all this is going on, the dense cores of both
galaxies will spiral in, as a result of dynamical friction, in the
central regions of the collision.  Finally they, too, merge to
form a single core.

Many, if not most, galaxies harbor a massive black hole in their
center.  Recently, many such black holes have been detected, with
masses spanning a range from a million to a billion solar masses, up
to $0.1$\% of the mass of the parent galaxy \cite{R999}.  When two galaxies
collide and stick together to form one large merger remnant, the dense
nuclei of the two parent galaxies will spiral in, within the central
region of the newly formed galaxy.  These nuclei will merge to form a
single dense nucleus, as soon as they come in contact with each other.
What will happen, if each nucleus contains a black hole, however is
far from clear.

At first, they will keep circling each other, within the single newly
formed dense nucleus.  Although dynamical friction tends to let them
spiral in rapidly at first, this process becomes considerably less
efficient by the time the amount of mass in stars between the two
holes becomes smaller than the mass of the holes themselves.  The
stars that initially tend to be most efficient in providing a braking
mechanism are scattered into different orbits.  As a result, the
system may reach a stagnation point, in which little further dynamical
friction occurs.

The prediction of this stagnation process was made almost twenty years
ago \cite{R10}, and since then many attempts have been made to check
this prediction quantitatively, using large-scale $N$-body
calculations.  Until the advent of the first GRAPEs, this problem was
completely intractable, even on the largest supercomputers available.
One reason that the GRAPE computers are suitable for this type of
problem is the intrinsically high dimensionality of the problem.  With
two black holes in an eccentric orbit around each other, there is no
symmetry in either configuration space or velocity space.  As a
result, the stellar dynamics problem is truly six-dimensional, when
seen as a fluid flow in phase space.

In contrast, modeling a globular cluster is often done by assuming
spherical symmetry, which leaves only one spatial dimension (radial)
and two velocity dimensions (radial and tangential) to worry about.
In practice, further simplifications have often been made, in which
the distribution function of the stars is assumed to be dependent only
on energy, or sometimes on energy and angular momentum.  Fokker-Planck
methods have therefore been very useful, initially, in modeling
globular clusters, especially during the core collapse phase.  After
core collapse, during the reexpansion phase, the effects of binaries
have to be taken into account, an extremely granular process that
defies the main Fokker-Planck assumptions of smoothness of the
distribution function.  However, even so, it has been very useful to
compare the full $N$-body calculations in the post-collapse domain
with approximate Fokker-Planck treatments.  However, a Fokker-Planck
treatment of a six-dimensional system is completely impractical from
the outset.

The first attempts to use the GRAPE to tackle this problem, were made
in 1990 \cite{R11}, using the GRAPE-2, followed by more recent attempts
\cite{R12} on the GRAPE-4.  Three
important conclusions have emerged from these studies.
(i) When two identical galaxies, each harboring a central black hole, merge,
they will produce a merger remnant with a ratio of core radius $r_c$
to half-mass radius $r_h$ that is comparable to that of the original
galaxies.  In contrast, galaxies without black holes tend to produce
merger remnants in which $r_c/r_h$ is smaller than in the original
galaxies.  In the former case, $r_c/r_h \sim
M_{BH}/M_{tot}$, where $M_{BH}$ is the mass of the central black hole,
and $M_{tot}$ is the mass of the whole galaxy.
(ii) This `core', formed around the black hole binary after the merging
of the two galaxies, does not have a completely flat density
distribution in the center.  In fact, it looks more like the `weak
cusps' observed in many galaxies by the Hubble Space Telescope \cite{R13}.
The formation mechanism of this cusp is not well understood.
(iii) Whether or not a black hole binary, lurking in the core of a merger
remnant, has had time to spiral in within the current age of the Universe,
and under which circumstances, is still largely an open question.
We expect the continuum limit to be reached for $N\simeq10^7$.
These calculations will only be feasible with the GRAPE-6 (Table 1).

\chapter{Formation and Evolution Processes, from Planets to Galaxy Clusters}

We will briefly discuss how the GRAPE computers have been used to
study the origin of structure in the Universe, from very small scales
to the largest scales that can be observed.  On the small end, the
coagulation of grains and boulders to form planets has been modeled,
in order to understand the formation process of our own planetary
system, as well as that around nearby stars.  Increasing the length
scale of interest by a factor of a billion, we discuss the formation
of galaxies.  Multiplying the size by another factor of a thousand,
we reach the scale at which rich clusters of galaxies evolve.

\section{Planet Formation}
    
After the Sun was formed, some matter of the proto-solar nebula was
left in a disk around the Sun.  Grains that condensed out of the
original gas coagulated through collisions to form larger and larger
particles, the size of pebbles, boulders, and larger proto-planetary
bodies.  To model this process in detail has turned out to be
difficult, because significant evolution takes place on a time scale
larger than a crossing time, by a factor of a million or more.

The main stumbling block has been the need to simultaneously model the
presence of a wide variety of particle sizes, or equivalently, masses.
A little more than ten years ago, it was realized that dynamical
friction plays an essential role in planetary
formation. \cite{Greenberg1978, WetherillStewart1989}  This process
forces more massive particles to have smaller random velocity, which
effectively increases their collision cross section. Thus, massive
particles can grow much more rapidly than less massive particles.

Kokubo and Ida \cite{R14} used the GRAPE-4 to model this type of growth of
planetesimals, under the assumptions that the accretion was perfect
({\it i.e.} the collisions were totally inelastic) and that there was
no gas left in the system to cause non-gravitational drag on the
particles.  They found the mass distribution to relax quickly to a
continuous power-law mass distribution with $dN/dm\propto m^{-2.5}$,
where $N$ is the cumulative number of bodies, independent of the
initial mass distribution (a result that was subsequently derived
analytically \cite{R15}).  Their most interesting result was that
the heaviest body would subsequently detach from the continuous power
law distribution, featuring a much more rapid growth in mass, called
runaway growth, that could lead to the formation of a planet.

Kokubo and Ida \cite{R16} again used the GRAPE-4 to study the later stages
of planet formation, on a more global scale.  The earlier local
run-away studies, leading to the formation of a single protoplanet,
give rise to multiple protoplanet formation when a large fraction of
the protoplanetary disk is modeled.  They found that such protoplanets
are formed and keep growing independently provided their orbital
separations are wide enough.  After a while, the growth rate of these
protoplanets slows down, because their gravitational perturbations
increase the random motion of the swarm of planetesimals they are
embedded in.  A continuous mass distribution of relatively light
planetesimals can thus coexists with a small number of large
protoplanets, for millions of years.

\section{Galaxy Formation}
    
To study the formation of a single galaxy, it is important to model
its environment, out to large distances, given the long-range
character of the gravitational force, which through tidal effects
influences the angular momentum distribution within the contracting
gas clouds destined to form galaxies.

In addition, it is essential to model the gasdynamical effects that
influence the early phases of galaxy formation.  While the
GRAPE has been designed primarily for stellar dynamical computations,
it has proved to be flexible in accommodating deviations
from an inverse square law.  A key property of the GRAPE hardware
is that it uses the inter-particle distances, that are computed in
order to calculate the pair-wise gravitational forces, to construct
for each particle, a list of neighboring particles that reside within
a prescribed distance.

Using this neighbor list, hydrodynamical simulations can be run on the
front end workstation.  The prime example here is smoothed particle
hydrodynamics, or SPH \cite{R18}.  Examples of these types of
simulations include the formation of galaxies \cite{R19}, the physical
origin of Ly-$\alpha$ and metal line absorption systems \cite{R20}, the
structure of galaxy clusters \cite{R21}, and the fragmentation of molecular
clouds \cite{R22}.

Simulations of galaxy formation have demonstrated that structure,
kinematics and chemical evolution of model galaxies which form in
hierarchical clustering scenarios agree with corresponding properties
of observed galaxy populations \cite{R23}. The major shortcoming is
that simulated galaxies are too concentrated. This is usually referred
to as the angular momentum problem \cite{R19} and suggests that
efficient feedback due to late stages of stellar evolution (for
example winds, and supernovae) is needed for a successful galaxy
formation model.

Simulations of damped Lyman-$\alpha$ absorption systems demonstrated
that non-equilibrium dynamics can easily explain the apparent
discrepancy between the observed high velocity of low ionization lines
and the relatively small circular velocity predicted by hierarchical
models of structure formation. \cite{R20} The evidence that
damped Lyman-$\alpha$ absorbers at high redshift are related to large
rapidly rotating disks, which would disagree with the hierarchical
clustering hypothesis, is thus not compelling. \cite{R24}

\section{Galaxy Cluster Evolution}

Galaxies are formed in a long drawn out process, starting somewhere
within the first billion years after the Big Bang, and continue to
form today.  Most galaxies are formed in isolation or in small groups,
but some galaxies are form in much richer groups, called clusters of
galaxies, or even superclusters of galaxies.  The typical properties
of galaxies formed in such clusters are different from galaxies
that were formed elsewhere.  For example, most galaxies in clusters
are elliptical galaxies, whereas most field galaxies are spiral
galaxies. \cite{BinneyMerri}

To what extent do these differences reflect the different formation
history of the galaxies, as they may have been affected by, for
example, the much higher matter density in the sites where rich
clusters of galaxies were born?  And to what extent do the differences
reflect later modifications to the galaxies, as a result of the
different dynamical environment of a rich cluster?  In attempts to
resolve this nature versus nurture debate, the GRAPE has been used to
model the internal evolution of a rich galaxy cluster.

Apart from the calculations by Bartelmann and Steinmetz \cite {R21},
already mentioned in the previous section, earlier work by
Funato {\it et al.} \cite{R25} simulated the evolution of clusters of
galaxies containing 32 to 128 galaxies. What they found is that `passive'
evolution of galaxies, caused by mutual encounters as well as by the
influence of the tidal field of the parent cluster, alters
the mass and size of individual galaxies. In particular, they
found that passive evolution leads to a distribution of masses with
$M(\sigma) \propto \sigma^4$, where $\sigma$ is the internal velocity
dispersion of the stars within a galaxy.

To understand the detailed mechanism of this passive evolution, Funato
and Makino \cite{R26} used the GRAPE-4 to study a large number of encounters
between two isolated galaxies, in order to determine how the resulting
changes of mass and binding energy depend on the models used for the
galaxies, and on the parameters describing the type of encounter.
They then estimated the cumulative effect of encounters, in the
setting of a rich cluster of galaxies.  They again found that the mass
distribution of galaxies tends to approach $M(\sigma) \propto
\sigma^4$, for the mass $M$ of a galaxy as a function of its velocity
dispersion $\sigma$.

Their results resembles the
observational Faber--Jackson relation, the empirical result that the
luminosity of a galaxy $L(\sigma) \propto \sigma^4$, for elliptical
galaxies.  Note that the remnants of collisions between galaxies
typically resemble elliptical galaxies, even if the progenitors were
spiral galaxies or other types of galaxies.  Because it is also
reasonable to assume that $M\propto L$, this agreement with
observations suggests that the encounters of galaxies play an
important role in the evolution of galaxies in a cluster of galaxies.


\bibliographystyle{nosort}

We thank Yoko Funato, Eiichiro Kokubo, and Matthias Steinmetz for
discussing their GRAPE results with us.  We also thank an anonymous
referee for helpful comments.

\end{document}